Electronic structure and ferromagnetic behavior in the $Mn_{1-x}A_xAs_{1-y}B_y$ alloys

A.V.Golovchan, I.F.Gribanov

*Donetsk Institute for Physics and Engineering named after O.O.Galkin of the National Academy of Sciences of Ukraine, 72, R.Luxemburg str. Donetsk 83114, Ukraine*

In this work, a systematic *ab initio* study of the influence of doping on electronic structure and local magnetic moments of ferromagnetic *MnAs* has been carried out. The majority of the considered substitution elements, potentially suitable for modification of *MnAs* as working material of magnetic refrigerators is shown to result in reduction of the ferromagnetic moment. But there are variants of anionic substitution (for example, substitution of As for S and Se) when the magnetic moment grows.

## 1. Introduction

Recently, the development of a new magnetic refrigeration technology, based upon the magnetocaloric effect (MCE), has brought an alternative to the conventional gas compression technique in a wide range of temperatures including the room ones. This is due to higher cooling efficiency and comparative environmental cleanliness of magnetic refrigerators. For magnetic refrigerators the *MnAs* is among the promising ones owing to giant magnetic entropy change, which accompanies the magnetic field-induced first-order phase transition from paramagnetic(PM) to ferromagnetic(FM) state near the Curie point $T_C=318K$ ($\Delta S \approx -40 J/kg \cdot K$ with the magnetic field $H$ varying from 0 to 5 T, the positive MCE) [1]. As known[2], in stoichiometric *MnAs* both spontaneous and magnetic field-induced *PM-FM* phase transitions occur with crystal structure change $B31 \rightarrow B8_1$ and are accompanied by a high (field or thermal) hystereses which is bad for practical use. The problem can be solved by a suitable cationic or anionic substitution providing the conservation or a small expansion of the crystal lattice (otherwise at low temperatures the *B31* crystal structure is stabilized resulting in the suppression of the «high-spin» ferromagnetic state and disappearance of the *PM-FM* phase transition we are interested in) [1,2]. The choice of the optimal substitution is not a trivial problem because it is necessary to satisfy several conditions at a time: the reduction of a hystereses of phase transition, preservation or increasing the magnetization jump with high speed of its change during the *PM-FM* phase transition, and the obtaining of such transition in the necessary temperature range. In searching for the optimal *MnAs* doping it is necessary to rely upon theoretical backgrounds which take changes in the electronic structure into account. The first-principle studies of the binary *MnAs* are numerous (see e.g [3,4] and references there in). However, the electronic structure of *MnAs*-based alloys with *NiAs*-type(*B8₁*) crystal structure has not been studied yet. This paper deals with a systematic *ab initio* study of the influence of anionic and cationic substitution and crystal lattice parameters variation on the electronic structure and local magnetic characteristics of ferromagnetic MnAs with crystal structure *B8₁*(NiAs-type).

## 2. Computational details

The calculations are carried out with the fully relativistic Korringer-Kohn-Rostoker - coherent potential approximation(KKR-CPA) method (the Munich SPRKKR package [5]) within the atomic-sphere approximation for crystal potential. The exchange-correlation potential is chosen in the local spin-density approximation with the parametrization given by Vosko, Wilk and Nusair [6]. In the calculations we used the experimental parameters of the hexagonal lattice as basic parameters (hexagonal *B8₁* crystal structure, symmetry group *P6₃/mmc*, $a = 3,730 \overset{o}{A}$, $c = 5,665 \overset{o}{A}$, $c/a = 1,5188$. The unit cell consists of two Mn atoms at the Wyckoff 2a sites (0, 0, 0) and (0, 0, 1/2) and two As atoms at the Wyckoff 2c sites (1/3, 2/3, 1/4) and (2/3, 1/3, 3/4). It also contains two empty spheres at the Wyckoff 2d sites (1/3, 2/3, 3/4) and (2/3, 1/3, 1/4)) [7]. Partial densities of electronic states corresponding to these parameters are shown in Fig.1. The bottom filled bands contain As 4s states. The conduction band is located above $E = 0,1 Ry$ and formed basically by Mn 3d states and As 4p states that point to a strong p-d hybridization in the system under study. The so-hybridized states are also present near the Fermi level. They are an important peculiarity of the like materials that influences the exchange splitting of *Mn 3d*-states. The electronic structure is, as a whole, typical of *Mn* pnictides and similar to those obtained in some other papers (see, for example, [8]).

The calculated magnetic moment of manganese atoms $M_{Mn}$=*3.36* $\mu_B$, the total magnetic moment per formula unit $M_{MnAs}$=*3.14* $\mu_B$. For comparison, in [8] $M_{Mn}$=*3.32* $\mu_B$, $M_{MnAs}$=*3.22* $\mu_B$ and the experiment (neutron diffraction investigation) gives $M_{MnAs} = 3.3 \pm 0.1 \mu_B$ for *T=4.2 K* [9]. As seen, in both cases the agreement with the experiment is quite satisfactory.

### 3. Results and discussion

### 3.1. Variation of the lattice constants

The influence of variation of the lattice constants on values of atomic and total magnetic moments of *MnAs* is shown in Fig.2. According to the results of other authors (see for example, [10]) there is the increase of $M_{Mn}$ and magnetization per unit cell $M_{total}$ with the obvious tendency to saturation at expansion of the lattice. The value of the saturation magnetic moment $M_{Mn} \approx 4\mu_B$ can be explained by charge transfer from *Mn* to *As* for filling the electronegative As 4p states. The remaining four valence electrons of Mn atoms form, according to Hund`s rules, the magnetic moment of 4 $\mu_B$. For basic lattice parameters there is only a partial charge transfer due to strong p-d interaction which also leads to a high negative polarization of anions ($-0,23 \pm 0,05 \mu_B$, a neutron diffraction investigation [11]). Lattice expansion decreases the degree of *p-d* hybridization and increases the ionicity of the system. Note that the results obtained for the compressed lattice are hypothetical because a *B8₁(FM)* phase becomes unstable already at small compression.

### 3.2. Cationic substitutions

Here we consider only 3d metals as the elements which substitute *Mn* atoms in *MnAs*. Besides, the case of cationic vacancy origination is analysed too. As known [2,12], at the doping of *MnAs* the majority of 3d-metals cause the compression of the crystal lattice and, as a consequence, loss of stability of the ferromagnetic phase $B8_1$, stabilization of the *B31* structure and magnetic phases with the «low-spin» state of *Mn* atoms. Experiments show that with approaching $B8_1$ phase boundaries(at doping or under external pressure application) the sharp growth of entropy change *ΔS* is observed at the magnetic field-induced first-order *PM-FM* phase transition which correlates with the increase in volume jump by almost an order of magnitude. So, for *P=2.23* kbar and magnetic field change of 5T the maximal entropy change $\Delta S \approx -270 \; J/kg \cdot K$, whereas for *P=0*, $\Delta S \approx -40 \; J/kg \cdot K$ [13]. Similarly, at substitution of 0.3 % *Mn* by *Cu* the entropy change increases up to $\Delta S \approx -170 \; J/kg \cdot K$ [14]. Such behaviour is accompanied by increase of spontaneous and magnetic field-induced *PM-FM* phase transition hystereses and is caused by strong interaction between magnetic and lattice subsystems. Large thermal and field hystereses present at first-order phase transition prevents the material from being used as a working medium in magnetic refrigerators.

As known, in *MnAs* reduction of the specified hystereses occurs at the crystal lattice expansion[1,2]. For considered alloys $Mn_{1-x}A_xAs$ (*A=3d* metal) the growth of the unit cell volume with *x* increase is only in the case of *A=Ti* [15]. Thus, the ferromagnetic phase with $B8_1$ structure is conserved for enough large *x*. For $Mn_{1-x}Ti_xAs$ we have performed calculations of the density of electronic states and magnetic characteristics at the basic lattice parameters. The calculated atomic moments and total magnetic moment per unit cell in $Mn_{1-x}A_xAs$ (*A=Ti,Cu,Vc*) are shown in Fig.3. Similar results for cases of substitution of *Mn* atoms by vacancies or Cu atoms are given there for comparison. It is seen that the increase of *Ti*, *Cu* and *Vc(Mn)* concentration leads to reduction of the total magnetic moment per unit cell, and the presence of *Ti* atoms does not practically affect the local magnetic moment of remained *Mn* atoms, whereas *Vc(Mn)* and *Cu* atoms suppress it noticeably and affect $M_{total}$ to a higher extent. In the two cases the spin polarization of *As* atoms behaves differently with *x* growth: decreasing for *A=Ti* and increasing for *A=Vc(Mn)*. As follows from Fig.3, in $Mn_{1-x}Ti_xAs$ the reduction of $M_{total}$ with $x$ increase is, mainly, due to that in the quantity of magnetic atoms in the crystal.

Similar calculations and comparison with experimental data were done for the $Mn_{1-x}Ti_xAs$ system with the experimental lattice parameters taken from literature[15]. The results are shown in Fig.4. In the bottom part the experimental dependences of lattice constants on $x$ are given. In the considered range of $x$ concentration $M_{total}$ decreases linearly with $x$ growth and for $x = 0.1$ $\Delta M_{total} = -0.6\mu_B$ ($\Delta M_{MnAs} = -0.3\mu_B$). In the case of doping experiments we have $\Delta M_{MnAs} = -0.4\mu_B$ (the magnetic measurements [2]). Comparison of Figs. 3 and 4 shows that qualitative distinction is only for the behaviour of spin polarization of $Ti$ atoms. Remaining very small, it increases with $x$ in the first case, and decreases in the second one. As a whole, the account of a variation of lattice parameters in this case does not affect significantly the considered magnetic characteristics.

Figure 5 illustrates the influence of the above cationic doping on spin-resolved total and site decomposed densities of the electronic states. It is obvious that $Cu$ states form a narrow band and lay (basically) below the Fermi level, whereas the band formed by $Ti$ states is much broader, it contributes to the total density of states at the Fermi level appreciably.

To sum up, we emphasize that substitution of $Mn$ atoms in $MnAs$ by atoms of 3d metals is characterized by the tendency when the total magnetic moment per unit cell decreases with dopant concentration increase.

### 3.3. Anionic substitutions

For the anionic substitution elements we took some elements of V and VI groups of the periodic table which, according to literature, can provide the above-mentioned conditions for the rational $MnAs$ doping. The influence of anionic substitution, for fixed(basic) lattice constants, on atomic magnetic moments and total magnetic moment per unit cell is shown in Fig.6. It is established that a part of the considered elements slightly decreases the magnetic moment of cation and total magnetization with $y$ increase (it concerns, in particular, $Sb$ chosen for the doping of $MnAs$ at experimental research of the MCE there[1]). However, $Se$ and $S$ substituting $As$ in $MnAs$, lead to essential growth of the magnetization. Even the greater growth is at the formation of vacancies in the As sublattice, but in practice such situation is apparently impossible, since the synthesis of *MnAs* with *Mn* excess will most likely result in precipitation of free metal at the grain boundaries and on sample surface. Other opportunity related to introduction of excess *Mn* atoms in interstitial sites of *NiAs* structure (by analogy with $Mn_{1+\delta}Sb$) is improbable because the volume of the mentioned interstitial sites is small. Nevertheless, such an opportunity has been analysed and it follows that the presence of excess *Mn* in the lattice leads to reduction of $M_{Mn}$ in regular cationic sites and of the magnetization per unit cell as well. Note also that the formation of vacancies in *As* sublattice should cause lattice compression and suppression of $B8_1(FM)$ phase. In addition, in all cases of the anionic doping the negative spin polarization of *As* atoms decreases.

In order to verify the used approach we have made calculations for isostructural $MnA_\delta Sb$, ($A=Cr,Mn,Fe,Co$) alloys with experimental lattice parameters [16]. For these alloys the neutron diffraction data are available**.** It is shown that excess cations occupy interstitial sites, and *Cr* and *Mn* atoms do not have magnetic moment, whereas magnetic moment of *Fe* and *Co* equals 0.84 $\mu_B$ [16]. The results of calculations and their comparison with the experiment are shown in Fig.7. Note that in all the cases the increase of the excess cation concentration $\delta$ increases parameter *a* and decreases *c* of the hexagonal *NiAs* lattice, so the unit cell volume grows a little, and *c/a* - falls. As seen, the introduction of excess atoms in interstitial sites decreases magnetic moments of *Mn* and *Sb* in the regular sites. In every case the magnetic moments of excess atoms are close to zero (the experimental situation for *A=Fe* and *Co* is not reproduced). The difference between the calculated local magnetic moments of impurity atoms and the experimental values is probably due to the fact that the intratomic Coulomb repulsion which in this case, apparently, plays an essential role[16] was not taken into account in calculations. Nevertheless, the calculated behaviour of the total magnetic moment per unit cell is in a good agreement with the experiment.

### 4. Conclusion

The systematic *ab initio* investigation of doping effect on electronic structure and some magnetic characteristics of *MnAs* has been done. Comparison of the obtained results with experimental data which are available in the literature shows that the method allows analyzing correctly enough the behaviour of magnetic moments in *MnAs*-based substitution alloys under doping and variation of the lattice parameters. The analysis has shown that the majority of dopants potentially suitable for the doping of *MnAs* as working material for magnetic refrigerators decrease the magnetization of ferromagnetic phase. Nevertheless, there are some cases of anionic substitution (for example *S, Se*), increasing it, that can be of practical interest.

The question of experimental realization of the specified variants demands special studies, as in literature there is a mention of small solubility of chalcogenides (*S, Se*) in *MnAs* [14].

**Acknowledgments**

We acknowledge financial support from Ukrainian FRSF(project №14.1/024).

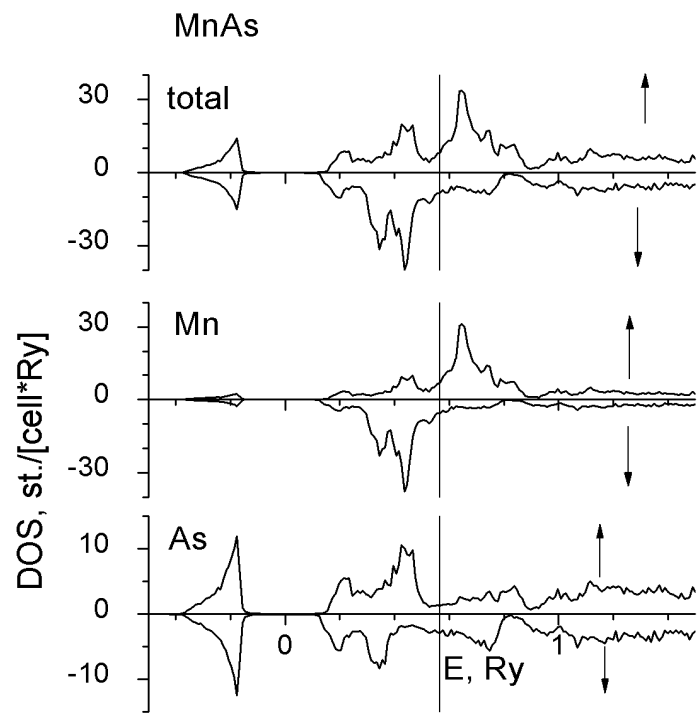

Figure 1. The spin-resolved total and site decomposed density of electronic states of *MnAs*.

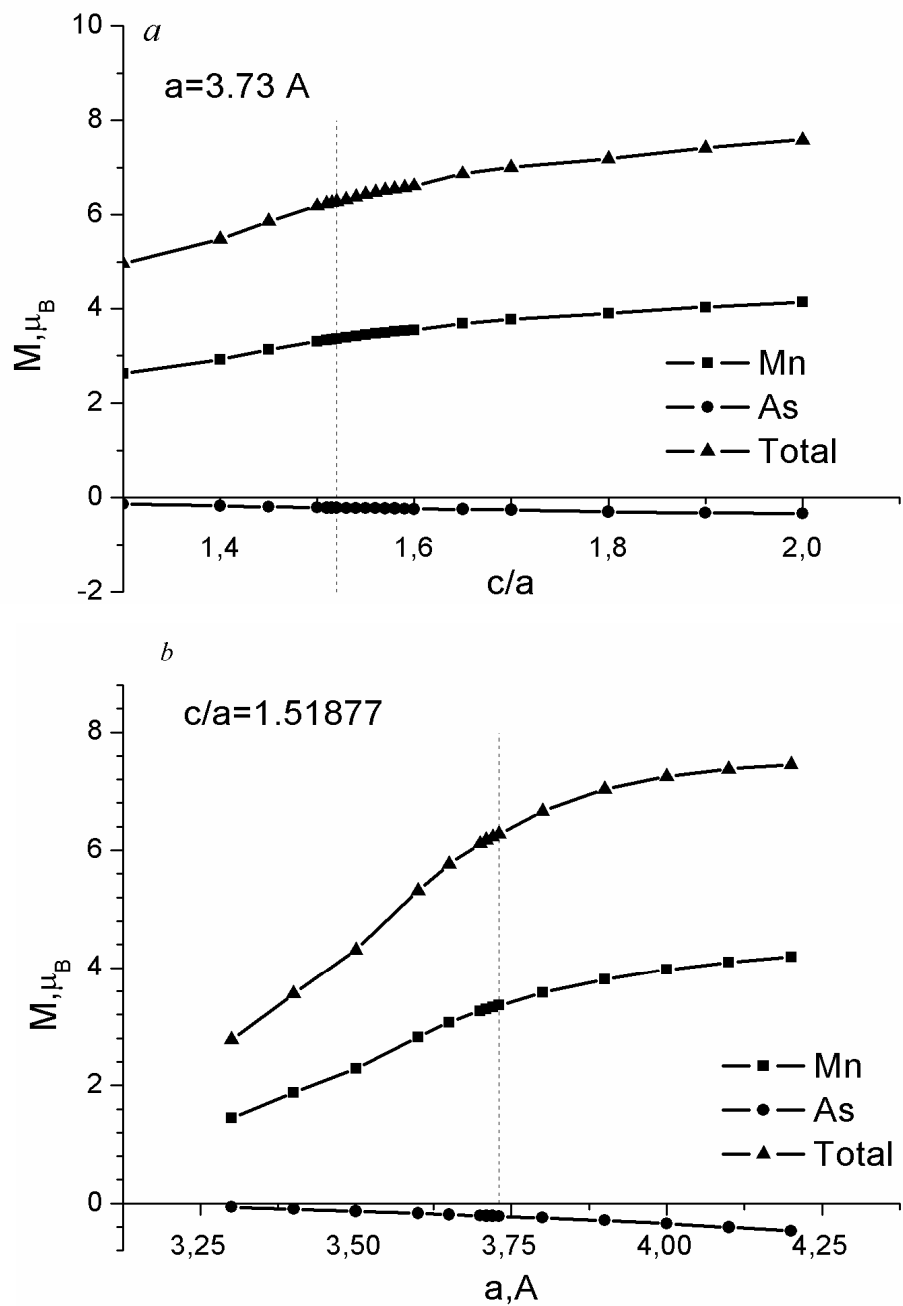

Figure 2. Magnetization, per unit cell(Total), and local magnetic moments for the compound *MnAs* as a function of lattice parameters.

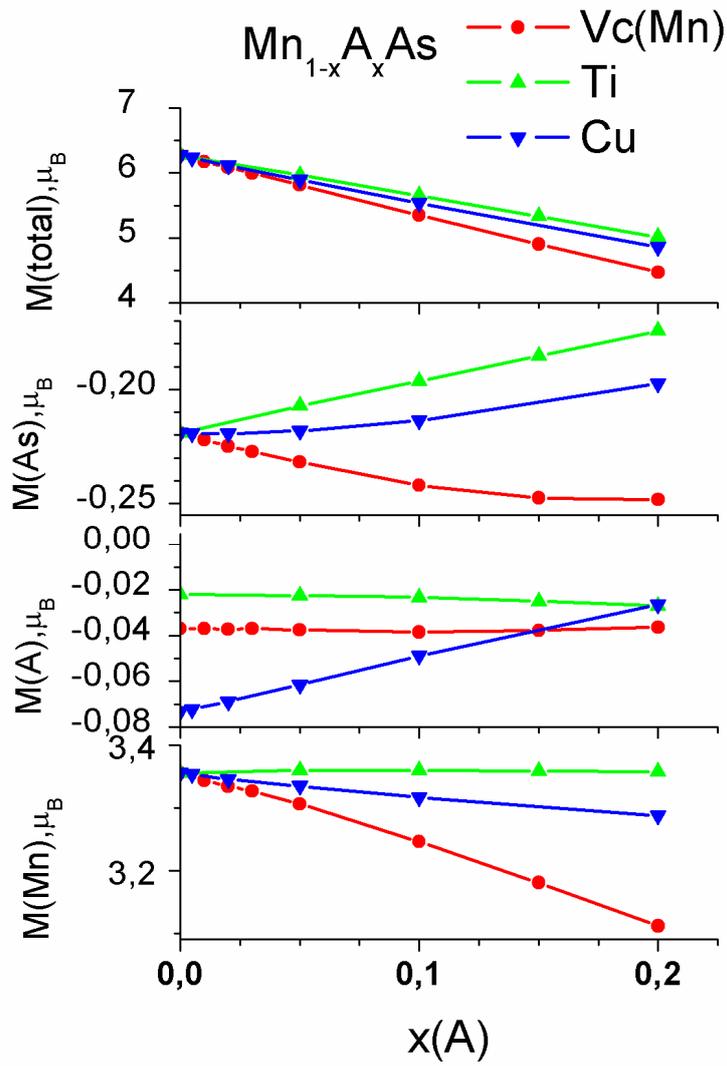

Figure 3. (color online) Influence of cationic substitution on local magnetic moments and $M_{total}$ of *MnAs* at fixed lattice parameters.

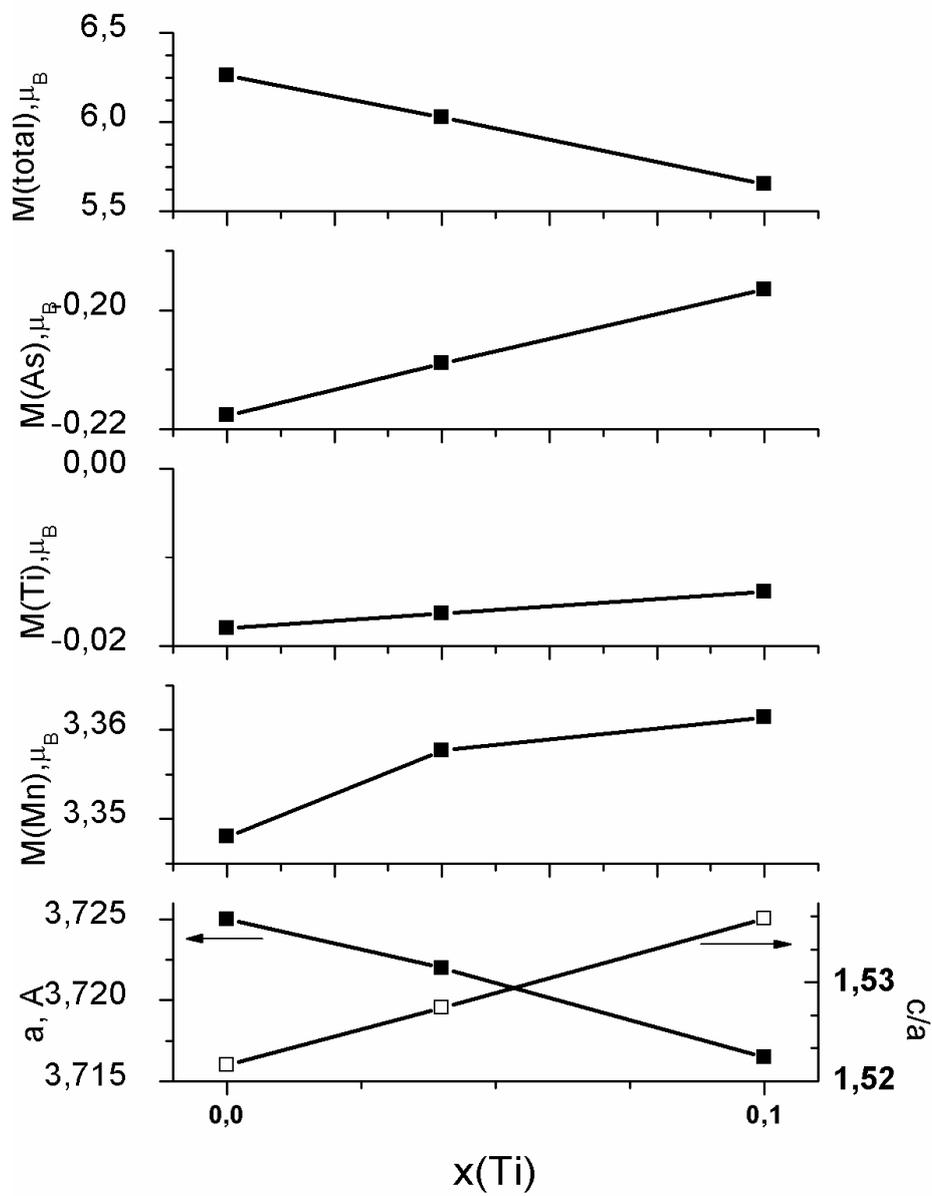

Figure 4. Calculated concentration dependences of local magnetic moments and $M_{total}$ in system $Mn_{1-x}Ti_xAs$ at the experimental lattice parameters.

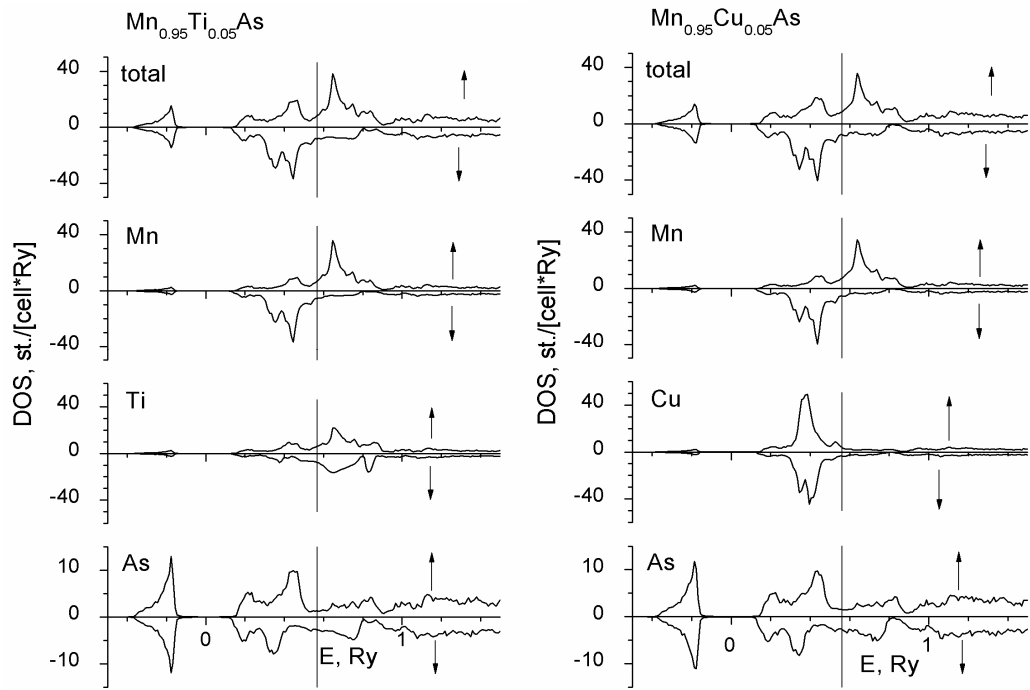

Figure 5. The spin-resolved total and site decomposed density of electronic states for
*Mn$_{0.95}$A$_{0.05}$As* (*A=Ti,Cu*).

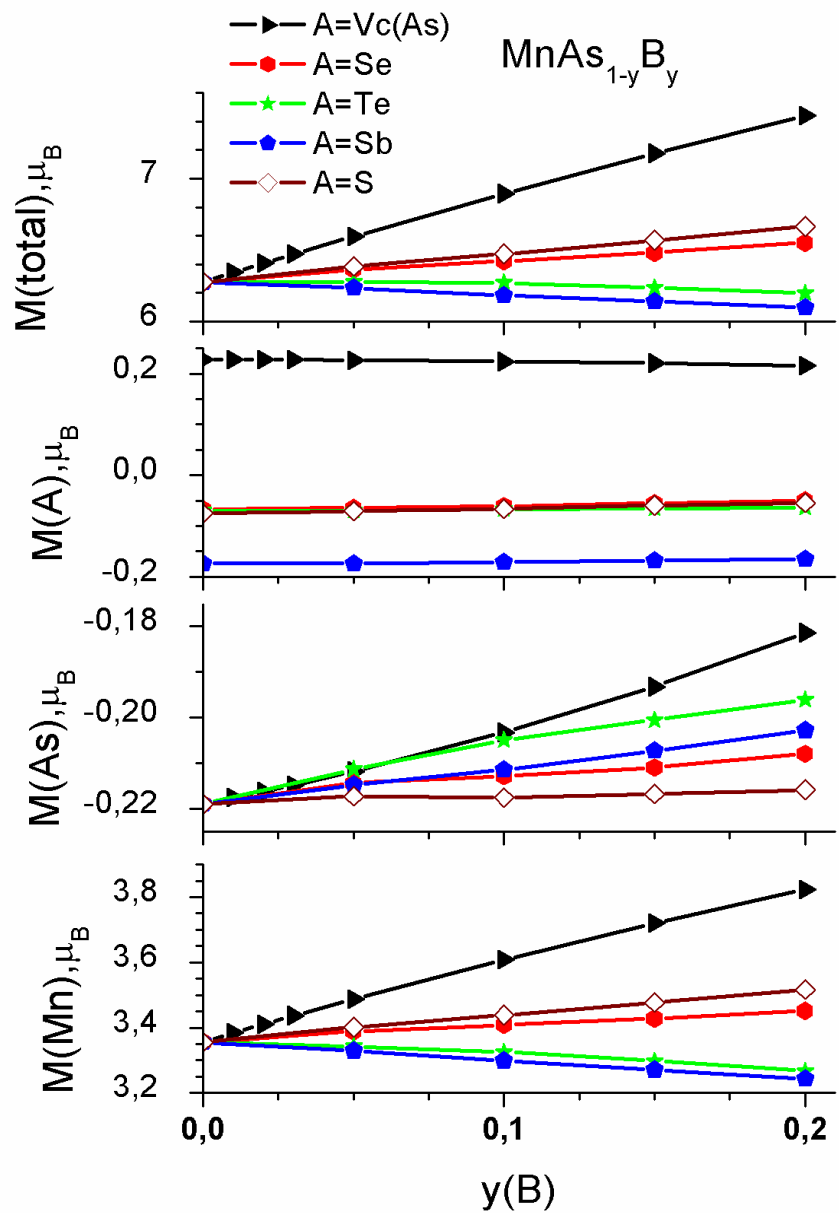

Figure 6. (color online) Influence of anionic substitution on local magnetic moments and $M_{total}$ of *MnAs* at fixed lattice parameters.

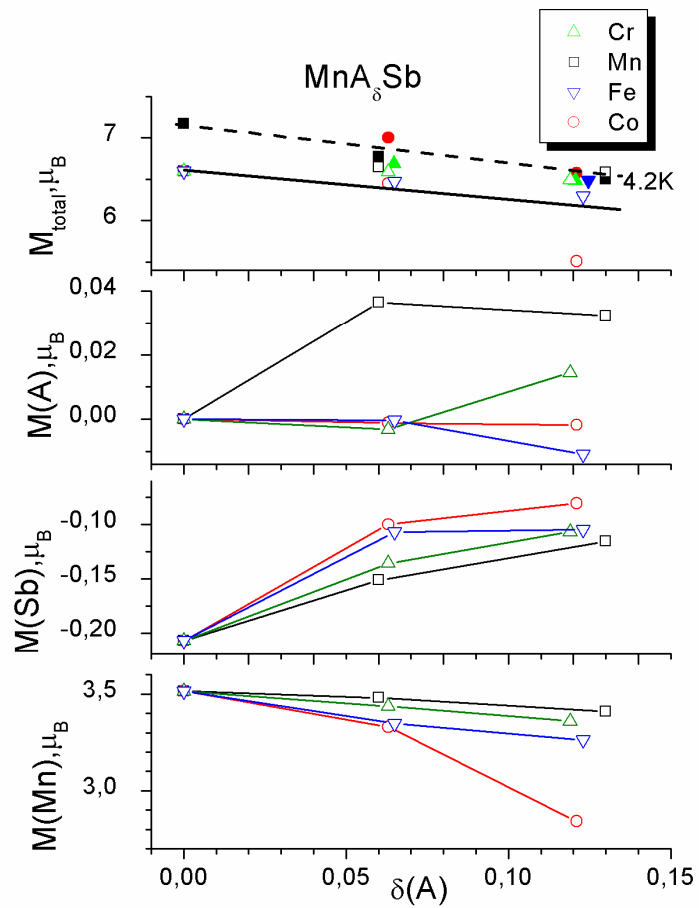

Figure 7. (color online) Component and concentration dependence of the magnetization per unit cell and local magnetic moments of *MnM$_\delta$Sb* (*M=Cr*, *Mn*, *Fe* and *Co*). Open symbols – the calculation, filled symbols and a dotted line(upper figure)– experiment[15].